\documentclass[hyper]{JHEP3}

\usepackage[dvips]{graphicx}
\usepackage{epsfig,multicol,amssymb,amsmath,cancel}

\newcommand\fverb{\setbox\fverbbox=\hbox\bgroup\verb}
\newcommand\fverbdo{\egroup\medskip\noindent%
            \fbox{\unhbox\fverbbox}\ }
\newcommand\fverbit{\egroup\item[\fbox{\unhbox\fverbbox}]}
\newbox\fverbbox

\def\bea{\begin{eqnarray}}
\def\eea{\end{eqnarray}}
\def\beq{\begin{equation}}
\def\eeq{\end{equation}}

\newcommand{\GeV}{\mathrm{GeV}}
\newcommand{\soft}{\mathrm{soft}}

\title{Light Higgsino in Heavy Gravitino Scenario with
Successful Electroweak Symmetry Breaking}

\author{Kwang Sik Jeong,
Masatoshi Shimosuka\footnote{
Current address: Ministry of Education, Culture, Sports, Science
and Technology (MEXT), Tokyo 100-8959, Japan},
Masahiro Yamaguchi \\
Department of Physics, Tohoku University, Sendai 980-8578, Japan  \\
E-mail :
\email{ksjeong@tuhep.phys.tohoku.ac.jp},
\email{shimosuka@tuhep.phys.tohoku.ac.jp},
\email{yama@tuhep.phys.tohoku.ac.jp} }

\preprint{TU-897}

\abstract{
We consider, in the context of the minimal supersymmetric standard model,
the case where the gravitino weighs $10^6$ GeV or more, which is preferred
by various cosmological difficulties associated with unstable gravitinos.
Despite the large Higgs mixing parameter $B$ together with the little hierarchy
to other soft supersymmetry breaking masses, a light higgsino with an
electroweak scale mass leads to successful electroweak symmetry breaking,
at the price of fine-tuning the higgsino mixing $\mu$ parameter.
Furthermore the light higgsinos produced at the decays of gravitinos can
constitute the dark matter of the universe.
The heavy squark mass spectrum of $O(10^4)$ GeV can increase the Higgs boson
mass to about 125 GeV or higher.
}

\keywords{Supersymmetry breaking, Supersymmetric Standard Model}

\begin{document}

\section{Introduction}

Although supersymmetry (SUSY) \cite{Nilles:1983ge,Haber:1984rc}
is a promising candidate for physics beyond the standard model (SM),
a closer look reveals its weak spots.
Among other things,  the gravitino with the very long lifetime is
known to be a potentially dangerous existence in  cosmology
\cite{Khlopov:1984pf,Kawasaki:1994af}.

The abundance of the unstable gravitinos is severely
constrained by the success of the big-bang nucleosynthesis for its
mass up to $\sim 5\times 10^4\: \mathrm{GeV}$
\cite{Kawasaki:2004yh,Kohri:2005wn, Kohri:2005ru}.
Even for a heavier gravitino, cosmology is not yet free from
the fear of the gravitino decay.
When the gravitinos were amply produced at the decay of the moduli
\cite{Endo:2006zj, Nakamura:2006uc, Dine:2006ii} (see also
ref. \cite{Joichi:1994ce} for an earlier discussion)
or other scalar fields \cite{Endo:2007ih,Endo:2007sz,Asaka:2006bv},
or in the thermal bath with high reheat temperature,
the lightest superparticles (LSPs) produced by the gravitino
decays would exceed the observed abundance of the dark matter
of the universe.
Given a neutralino LSP with mass around $100\: \GeV$, the
gravitino should weigh $10^6\: \GeV$ or even more
\cite{Nakamura:2006uc}.
To solve this problem, a previous work postulated the existence
of a lighter LSP in an extension of the minimal supersymmetric
standard model (MSSM) \cite{Nakamura:2008ey}.

In this paper, we shall revisit this problem within the framework of the MSSM.
The gravitino mass is generically related to soft SUSY breaking masses.
This is because the vacuum expectation value of the chiral
compensator auxiliary field $F_\phi$ is comparable to the gravitino
mass $m_{3/2}$ in size unless one considers a very specific SUSY
breaking scenario.
In this case, the SUSY-breaking Higgs mixing parameter $B$ is
comparable to $F_\phi$, and thus to $m_{3/2}$.
Furthermore, we assume that the contribution from the anomaly
mediation to other soft masses is inevitable and is not cancelled by other
SUSY breaking mediation contributions.
Thus the soft masses should satisfy
\begin{align}
m_{\soft} \gtrsim \frac{1}{8\pi^2}m_{3/2}.
\end{align}
On the other hand, the supersymmetric higgsino mass parameter
$\mu$ is not constrained by this argument.
We are thus led to consider the case where the higgsino is
light with the hierarchical mass spectrum\footnote
{Another possibility one can consider is that a
relatively heavy neutralino will annihilate very effectively
via Higgs resonance to reduce its relic abundance.
However, we will discard this case as it requires
fine turning of the masses of the neutralino and the Higgs boson.}:
\begin{align} \label{eq.mass.spectrum}
\mu \sim \mathcal{O}(100)\: \GeV \ll m_{\soft} \sim
\mathcal{O}(10^4)\: \GeV \ll m_{3/2} \sim \mathcal{O}(10^6)\: \GeV.
\end{align}

We note that the suppressed soft masses compared to the gravitino
mass are realized in the KKLT setup \cite{Kachru:2003aw}.
In this case, the resulting soft masses are of the mixed modulus-anomaly
mediation \cite{Choi:2004sx,Choi:2005ge,Endo:2005uy,Choi:2005uz,Falkowski:2005ck}.
Our argument given here can apply to a wider class of models and thus
we do not specify a particular mediation mechanism.

As we will show shortly, the electroweak symmetry breaking (EWSB)
successfully takes place with this hierarchy.  Furthermore the higgsino
LSP abundance produced by the gravitino decays is consistent with the
measured value of the dark matter abundance
and hence can constitute the dark matter of the universe.

The Higgs sector in this scenario contains a SM-like Higgs boson whereas
all others become very heavy $\sim O(10^4)$ GeV. We will show that
the SM-like Higgs boson naturally has mass in the range suggested by
the recent data from ATLAS and CMS at the Large Hadron Collider (LHC)
\cite{Higgs-LHC}.

\section{Electroweak Symmetry Breaking}

Let us begin by investigating how the EWSB takes place with
the mass spectrum eq.~\eqref{eq.mass.spectrum}.
The neutral part of the tree-level Higgs potential in the
MSSM is given by
\begin{align} \label{eq.reduced.scalar.potential}
V =& (|\mu|^2 + m_{H_u}^2)|H_u^0|^2 + (|\mu|^2 +
m_{H_d}^2)|H_d^0|^2 -(B\mu H_u^0H_d^0 + c.c.)
\notag \\
& +\frac{1}{8}(g_1^2 + g_2^2)(|H_u^0|^2 - |H_d^0|^2)^2 ,
\end{align}
where $m_{H_u}^2$ and $m_{H_d}^2$ are SUSY breaking mass squared parameters
for hypercharge $1/2$ and $-1/2$ Higgs doublet, respectively,
and $g_1 ,g_2$ are $U(1)_Y ,SU(2)_L$ coupling constants.

\begin{figure}[t]
\begin{center}
\begin{minipage}{15cm}
\centerline{
{\hspace*{0cm}\epsfig{figure=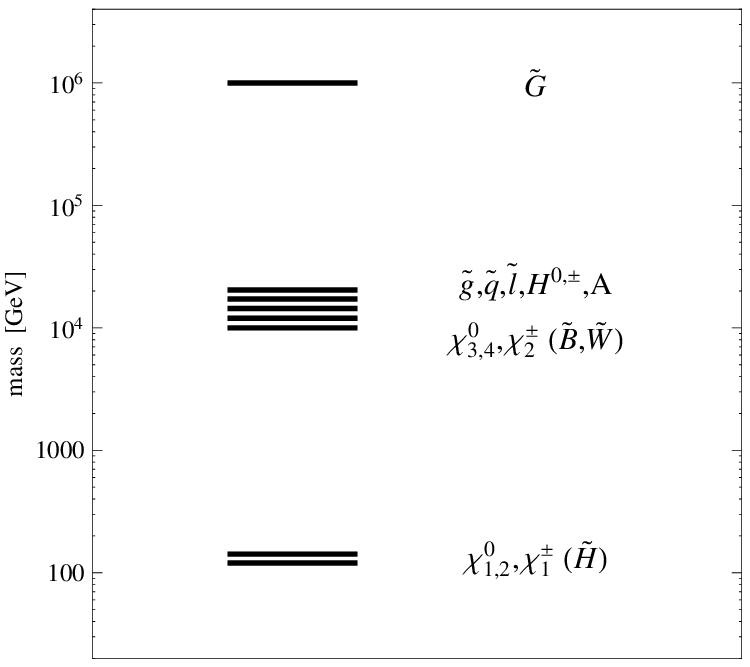,angle=0,width=7.cm}}
{\hspace*{.8cm}\epsfig{figure=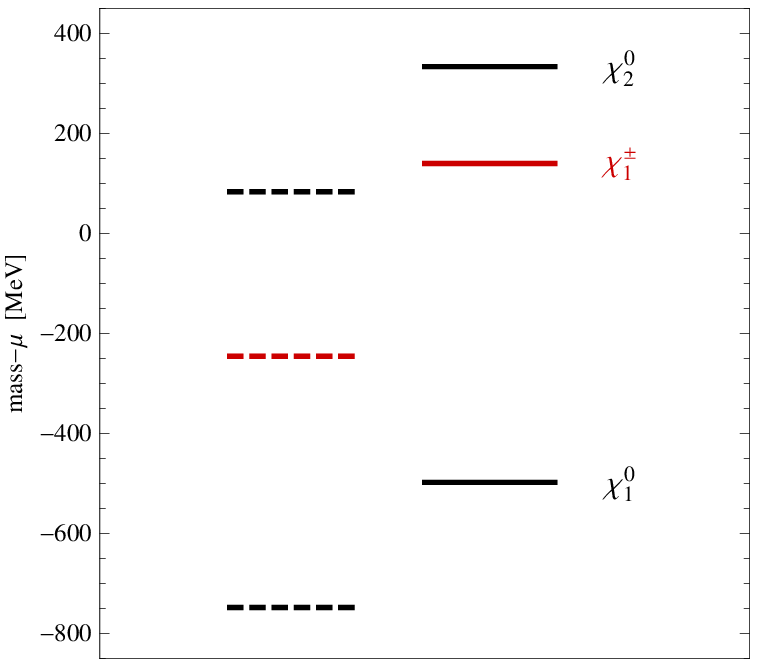,angle=0,width=7.cm}}
}
\caption{
The mass spectrum in heavy gravitino scenario with a light higgsino.
The right plot shows the masses of $\chi^0_{1,2}$ and $\chi^\pm_1$
for $\tan\beta=2$ (dashed lines) and $\tan\beta=10$ (solid lines) in
the case with $\mu=120\: \GeV$ and $M_1=M_2=10^4\: \GeV$.
}
\label{fig:spectrum}
\end{minipage}
\end{center}
\end{figure}

It is well known that the following two conditions should be
satisfied in order that the theory exhibits the EWSB:
\begin{enumerate}
\item The scalar potential is stable along the $D$-flat
direction, $|H_u^0|^2 =|H_d^0|^2$,
where the quartic terms are absent.
This yields
\begin{align} \label{eq.positive.Tr}
2|B\mu | < 2|\mu|^2 + m_{H_d}^2 + m_{H_u}^2 .
\end{align}
\item One of the eigenvalues of the squared-mass matrix is
negative, and so is the determinant
\begin{align} \label{eq.negative.det}
(|\mu|^2 + m_{H_d}^2)(|\mu|^2 + m_{H_u}^2)- |B\mu |^2 <0 .
\end{align}
\end{enumerate}

To realize the electroweak scale from the much larger soft
masses, the negative eigenvalue should be tuned to be at the electroweak
scale.
Then the magnitude of the determinant becomes much smaller
than the typical soft mass scale, which implies
$|B\mu|^2\sim m^2_{H_u}m^2_{H_d}$ for $\mu\ll m_{\rm soft}$.
The EWSB thus requires both $m^2_{H_u}$ and $m^2_{H_d}$
to be positive, and is driven by a large $B\mu$ term.
To be more precise, the extremum conditions of the Higgs potential
read
\begin{align}
m^2_{H_u} \simeq\,& \frac{m^2_{H_d}}{\tan^2\beta},
\notag \\
|\mu| \simeq\,& \frac{m^2_{H_d}}{|B|\tan\beta},
\end{align}
for $1\ll\tan^2\beta\ll m^2_{H_d}/|\mu|^2$ where
$\tan\beta=\langle |H^0_u| \rangle/\langle |H^0_d| \rangle $.
It is interesting to observe that
\begin{align}
\mu \sim \frac{1}{(8\pi^2)^2} m_{3/2},
\end{align}
when $B \sim m_{3/2}$ and $\sqrt{m_{H_u}m_{H_d}} \sim m_{3/2}/8\pi^2$,
which is anticipated from the anomaly mediation contribution.
For $m_{3/2} \sim 10^6\: \GeV$, one naturally obtains $\mu \sim 100\: \GeV$.
A rather small $m_{H_u}$ ($\sim m_{H_d}/\tan\beta$) at a low energy scale
would be the result of a negative radiative correction associated with
the top Yukawa coupling.

In fig.~\ref{fig:spectrum}, we illustrate the superparticle
mass spectrum in heavy gravitino scenario.
Since $\mu\ll m_{\rm soft}$, the lightest chargino and the two
lightest neutralinos are dominated by the higgsino components
and closely degenerate in masses.

\section{Higgs Boson Mass}

Below the scale $m_{\rm soft}\sim 10^4\: \GeV$, one combination of
$H_u$ and $H_d$ behaves like the SM Higgs doublet scalar while
sfermions, gauginos and other heavy Higgs bosons decouple from the theory.
In the low energy effective theory, the mass of the SM-like Higgs boson
$h$ can be estimated from the relation $m^2_h=\lambda(m_h) v^2$ where
$\lambda(m_h)$ is the Higgs quartic coupling renormalized at $m_h$,
and $v$ is the vacuum expectation value of the SM-like Higgs scalar.
The Higgs coupling $\lambda$ at $m_{\rm soft}$ is given by
\begin{align}
\lambda(m_{\rm soft}) = \frac{g^2_1+g^2_2}{4}\cos^2 2\beta
+ \frac{3y^4_t}{8\pi^2}\left(X^2_t-\frac{X^4_t}{12}\right),
\end{align}
and its low energy value is determined by the renormalization
group equations, which are affected by the higgsinos
with mass $\mu\sim 10^2\:\GeV$.
Here $y_t$ is the top Yukawa coupling, and $X_t=(A_t-\mu\cot\beta)/m_{\rm soft}$
is the stop mixing parameter.

\begin{figure}[t]
\begin{center}
\begin{minipage}{15cm}
\centerline{
{\hspace*{0cm}\epsfig{figure=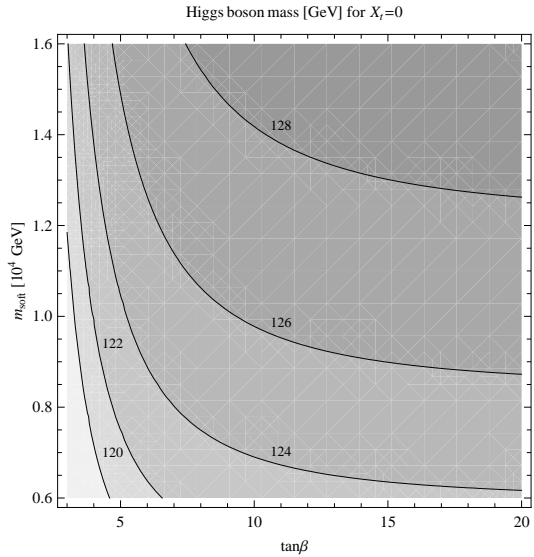,angle=0,width=7.cm}}
%{\hspace*{.8cm}\epsfig{figure=Fig_DM.eps,angle=0,width=7.cm}}
}
\caption{
The contours for the Higgs boson mass in the $\tan\beta$-$m_{\rm soft}$
plane for heavy gravitino scenario with $\mu=120\: \GeV$.
The stop mixing $X_t$ is set to zero.
}
\label{fig:mh}
\end{minipage}
\end{center}
\end{figure}

The Higgs quartic coupling $\lambda(m_h)$ gets a large positive contribution
from the loops involving the top Yukawa coupling
\cite{Okada:1990vk,Ellis:1990nz,Haber:1990aw}.
For heavy stops with mass $\sim 10^4\: \GeV$, this loop contribution
makes the Higgs boson have a mass around $125\: \GeV$, which lies in the
range where the LHC experiments reported an excess of Higgs-like events over
the background expectation \cite{Higgs-LHC}.
A stop mixing due to the trilinear $A$-term can raise the Higgs mass further.

Fig.~\ref{fig:mh} shows the Higgs boson mass for $\mu=120\: \GeV$ without
the stop mixing $X_t=0$.
One can see that $m_h$ is around $125\:\GeV$ for $m_{\rm soft}=10^4\: \GeV$.
The Higgs mass has a negligible dependence on $\mu$ since the higgsinos
only contribute to the beta function coefficients for $g_{1,2}$.
For instance, if the higgsinos decouple with mass $\mu\sim m_{\rm soft}$,
$m_h$ increases slightly by about 0.1\% compared to the case with
$\mu=120\: \GeV$.

\section{Higgsino Relic Abundance}

Let us next see whether the neutral higgsino is the LSP in the mass
spectrum eq.~\eqref{eq.mass.spectrum}.
We define the mass difference $\Delta m$ between the lightest
chargino $\chi^+_1$ and the lightest neutralino $\chi^0_1$ as
\begin{align}
\Delta m & \equiv  m_{\chi^+_1} -m_{\chi^0_1}
\notag \\
&= \Delta m^{(0)} +\Delta m^{(1)}_{\mathrm{gauge}}
+ \Delta m^{(1)}_{\mathrm{Yukawa}},
\end{align}
where $\Delta m^{(1)}$ is the 1-loop correction to the tree-level
mass difference $\Delta m^{(0)}$.
In the limit of $m_Z,|\mu| \ll M_1, M_2$ where $m_Z$ is the
$Z$ boson mass and $M_1,M_2$ are $U(1)_Y$ and $SU(2)_L$ gaugino masses,
the lightest chargino and neutralino consist mainly of the
charged and neutral higgsino, respectively.
Taking a rather unusual convention that $\mu$ is positive
while $M_1,M_2$ have either sign,
we find the two neutral higgsinos $\widetilde{H}_S,
\widetilde{H}_A$ have masses
\begin{align}
M_{\widetilde{H}_S} =\mu +\frac{1-\sin 2\beta}{2} m_Z^2
\bigg( \frac{\sin^2 \theta_W}{M_1} +\frac{\cos^2 \theta_W}{M_2}\bigg),
\notag \\
M_{\widetilde{H}_A} =\mu -\frac{1+\sin 2\beta}{2} m_Z^2
\bigg( \frac{\sin^2 \theta_W}{M_1} +\frac{\cos^2 \theta_W}{M_2} \bigg),
\end{align}
where $\theta_W$ is the Weinberg angle.
The charged higgsino has mass
\begin{align}
M_{\widetilde{H}^\pm} =\mu - \frac{m_W^2 \sin 2\beta}{M_2} ,
\end{align}
where $m_W$ is the $W$ boson mass.
Thus, the tree-level mass difference is
\begin{align}
\Delta m^{(0)} = M_{\widetilde{H}^\pm} -\min \left(
M_{\widetilde{H}_S}, M_{\widetilde{H}_A} \right).
\end{align}
We find that when $M_1$ and $M_2$ have the same sign, $\Delta
m^{(0)}$ is always positive.
But more generally, $\Delta m^{(0)}$ can take either sign.
The magnitude of $\Delta m^{(0)}$ is typically
\begin{align}
\left| \Delta m^{(0)} \right| \simeq 0.3 \:\GeV
\times \bigg(\frac{|M_2|}{10^4\: \GeV} \bigg)^{-1}
\left|1 + \frac{M_2}{M_1}\tan^2\theta_W \right|
\end{align}
for $\tan \beta \gg 1$.
Furthermore, the 1-loop correction from gauge boson loops
\begin{align}
\Delta m^{(1)}_{\mathrm{gauge}}
&=
\frac{g^2_2\sin^2\theta_W}{8\pi^2}
%\frac{\alpha}{2\pi}
|\mu | \int_0^1 dx (x+1)
\ln\left(1
+ \frac{1-x}{x^2}\frac{m^2_Z}{\mu^2}\right)
\notag \\
&= 0.24\: \GeV \to 0.35\: \GeV \quad ({\rm for}\, |\mu |=100\: \GeV
\to \infty )
\end{align}
is always positive, whereas that from the top-stop loops is
estimated as
\begin{align} \label{eq.top-stop}
\Delta m^{(1)}_{\text{top-stop}}
\sim -0.027 \: \GeV \times \bigg( \frac{m_t}{170\: \GeV}\bigg)^4
\bigg( \frac{m_{\tilde t}}{10^4 \: \GeV} \bigg)^{-2}
\frac{A_t}{10^4 \: \GeV}
\frac{1}{\sin^2 \beta},
\end{align}
where $m_t$ is the top mass, $m_{\tilde t}$ is the stop mass scale, and
$A_t$ is the mass scale included in the trilinear scalar term
\cite{Mizuta:1992ja}.
We represent the top-stop loop contribution explicitly
because it is the largest contribution in the Yukawa one.
As we can see in eq.~\eqref{eq.top-stop}, the Yukawa
contributions to the mass difference are negligibly small
compared to the gauge boson contribution.
Thus, we find
\begin{align}
\Delta m >0
\end{align}
in the sizable region of the parameter space.

Now we will discuss the relic abundance of the neutral higgsino LSP.
The LSPs are produced by the decays of gravitinos followed by
the pair-annihilation among them.
The abundance highly depends on the thermal averaged annihilation cross
section of the LSPs.
Here only the W boson or Z boson pairs are taken account
of in the final state because other annihilation processes are highly suppressed
due to the heavy soft masses.
Thus the cross section is computed to be \cite{Olive:1990qm}
\begin{align} \label{eq.cross.section}
\langle \sigma_{\mathrm{ann}} v_{\mathrm{rel}} \rangle
= \frac{g^4}{32\pi} \frac{1}{m_{\chi^0_1}^2}
\bigg[ \frac{(1-x_W)^{3/2}}{(2-x_W)^2} +\frac{1}{2\cos^4
\theta_W} \frac{(1-x_Z)^{3/2}}{(2-x_Z)^2} \bigg],
\end{align}
where $m_{\chi^0_1}\simeq |\mu|$, $x_W =m_W^2 /m_{\chi^0_1}^2$
and $x_Z =m_Z^2 /m_{\chi^0_1}^2$.

In computing the annihilation cross section, the Boltzmann distribution
was, for simplicity, assumed for the higgsino momentum distribution,
whose justification may be quite non-trivial \cite{Kawasaki:1995cy}.
We note, however, this simplification does not cause any significant
change to our result as far as the higgsinos are non-relativistic
and the annihilation occurs in the unsuppressed S-wave.
Also, we ignored the possible coannihilation effects \cite{Mizuta:1992qp},
which do not bring any dramatic alternation of our result.

\begin{figure}[t]
\begin{center}
\begin{minipage}{15cm}
\centerline{
{\hspace*{0cm}\epsfig{figure=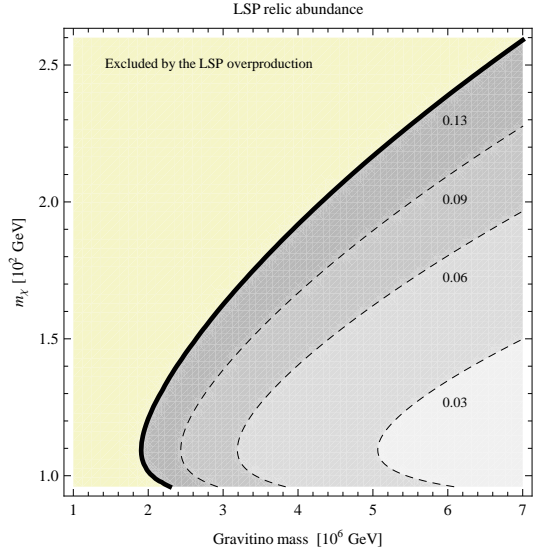,angle=0,width=7.cm}}
%{\hspace*{.8cm}\epsfig{figure=Fig_DM.eps,angle=0,width=7.cm}}
}
\caption{
The constant contours for the density parameter $\Omega_{\chi^0_1}h^2$ are
shown in the $m_{\chi^0_1}$-$m_{3/2}$ plane.
Tree dashed lines represent the contours of
$\Omega_{\chi^0_1}h^2 =0.03,0.06,0.09$, from the above respectively.
The real one stands for $\Omega_{\chi^0_1}h^2=0.13$
that is the 95 \% C.L. upper bound of the LSP abundance
restricted by the dark matter observation.
}
\label{fig:DM}
\end{minipage}
\end{center}
\end{figure}

As the nature of the $R$-odd particles, the gravitino decay yields
(at least) one LSP production under the $R$-parity conservation.
Thus in the absence of the annihilation among the LSPs, the final yield
of the LSPs would be the same as the initial yield of the gravitinos.
When the gravitino abundance is large enough, which we assume to be
the case, the annihilation among the LSPs becomes effective.
We note that the gravitino decay width is given by
\begin{align}
\Gamma_{3/2} = \frac{193}{384\pi}\frac{m^3_{3/2}}{M^2_{Pl}},
\end{align}
corresponding to the temperature at the gravitino decay
\begin{align}
T_{3/2} \simeq\,& \left(\frac{90}{\pi^2g_\ast(T_{3/2})}\right)^{1/4}
\sqrt{\Gamma_{3/2}M_{Pl}}
\notag \\
\simeq\,&
0.25 \GeV\times
\left(\frac{g_\ast(T_{3/2})}{10}\right)^{-1/4}
\left(\frac{m_{3/2}}{10^6\GeV}\right)^{3/2},
\end{align}
where the effective number of relativistic degrees of freedom $g_\ast$
changes from about 100 to 10 by the QCD phase transition at around 0.2 GeV.
Thus, the LSPs are produced after the freeze-out of the LSPs from the
thermal bath takes place (with the temperature
$T_f\simeq m_{\chi^0_1}/25-m_{\chi^0_1}/20$).
In this case, one can estimate the yield of the LSPs as \cite{Moroi:1994rs}:
\begin{align} \label{eq.yield.with.annihilation}
\frac{n_{\chi^0}}{s} \bigg|_{T_{3/2}}
\simeq \frac{H(T)}{\langle \sigma_{\mathrm{ann}}
v_{\mathrm{rel}} \rangle s} \bigg|_{T_{3/2}}
= \frac{1}{4} \left( \frac{90}{{\pi}^2 g_\ast (T_{3/2})}\right)^{1/2}
\frac{1}{\langle \sigma_{\mathrm{ann}}
v_{\mathrm{rel}} \rangle T_{3/2} M_{Pl}}.
\end{align}
From eq.~\eqref{eq.cross.section} and
eq.~\eqref{eq.yield.with.annihilation}, the ratio of the LSP mass
density to the entropy density is straightforwardly given by
\begin{align}
\frac{\rho_{\chi ^0_1}}{s}
\simeq\,\, & 0.22 \times 10^{-9} \: \GeV \nonumber \\
& \times \bigg[ \frac{1}{2\cos^4 \theta_W}
\frac{(1-x_Z)^{3/2}}{(2-x_Z)^2}+\frac{(1-x_W)^{3/2}}{(2-x_W)^2}
\bigg]^{-1}
\left( \frac{m_{\chi^0_1}}{100\GeV} \right)^3  \notag \\
& \times \left(\frac{g_{\ast}(T_{3/2})}{10}\right)^{-1/4}
\left( \frac{m_{3/2}}{10^6\GeV} \right)^{-3/2} ,
\end{align}
which corresponds to the density parameter
\begin{align}
\Omega_{\chi^0_1} h^2
\simeq  0.060
& \times \bigg[ \frac{1}{2\cos^4 \theta_W}
\frac{(1-x_Z)^{3/2}}{(2-x_Z)^2}+\frac{(1-x_W)^{3/2}}{(2-x_W)^2}
\bigg]^{-1}
\left( \frac{m_{\chi^0_1}}{100\GeV} \right)^3  \notag \\
& \times \left(\frac{g_{\ast}(T_{3/2})}{10}\right)^{-1/4}
\left( \frac{m_{3/2}}{10^6\GeV} \right)^{-3/2} ,
\end{align}
where $h \simeq 0.72$ is the Hubble constant in unit of 100~km/s/Mpc.

In fig.~\ref{fig:DM}, we draw constant contours for
the density parameter $\Omega_{\chi^0_1}h^2$ in the $m_{\chi^0_1}$-$m_{3/2}$
plane.
The real line shows $\Omega_{\chi^0_1}h^2 =0.13$ that is the 95 \% C.L. upper
bound of the LSP abundance restricted by
the dark matter observation \cite{Spergel:2003cb}.
Remarkably, there is no LSP overproduction problem in the
mass spectrum eq.~\eqref{eq.mass.spectrum}
and we can conclude the neutral higgsino constitutes the dark
matter of the universe.

\section{Discussion and Summary}

Since $\chi^0_{1,2}$ and $\chi^+_1$ are all higgsino-like for $\mu\ll m_{\rm soft}$,
only the interactions $Z\chi^0_1\chi^0_2$, $W^-\chi^+_1 \chi^0_{1,2}$,
$\gamma\chi^+_1\chi^-_1$ and $Z\chi^+_1\chi^-_1$ have non-negligible couplings.
Heavy gravitino scenario with the mass spectrum eq.~(\ref{eq.mass.spectrum})
generally leads to $m_\pi < \Delta m \lesssim 1\: \GeV$ with $m_\pi$ being
the pion mass, for which the lightest chargino decay is dominated by the single
pion mode $\chi^+_1\to\chi^0_1\pi^+$:
\begin{align}
\Gamma_{\chi^+_1\to\chi^0_1\pi^+} =\,& \frac{G^2_F}{\pi} \cos^2\theta_C
f^2_\pi \Delta m^3 \left(1-\frac{m^2_\pi}{\Delta m^2}\right)^{1/2}
\notag \\
\simeq\,&
\frac{1}{0.2{\rm cm}}\left(\frac{\Delta m}{500{\rm MeV}}\right)^3
\left(1-\frac{m^2_\pi}{\Delta m^2}\right)^{1/2},
\end{align}
where $f_\pi$ is the pion decay constant, and $\theta_C$ is the Cabbibo angle.
Thus, it would be difficult for the lightest chargino to produce a visible track
in the detector unless it is highly boosted.\footnote{
See refs.
\cite{Chen:1996ap,Cirelli:2005uq,Cheung:2005pv,Cirelli:2009uv,Buckley:2009kv,Hall:2011jd}
for discussions of collider searches.
}
Also, produced pions would be too soft to be detected.
The mass difference between the two lightest neutralinos is
similar to $\Delta m$ in size, and thus the detection of the decay products
of $\chi^0_2$ would be challenging as well.
On the other hand, at $e^+e^-$ colliders, the processes
$e^+e^-\to \gamma \chi^0_1\chi^0_2$, $\gamma \chi^+_1\chi^-_1$ mediated
by virtual $Z$ exchange become important, and would provide a visible signal
if a hard photon radiation occurs in the initial state.

The direct detection of dark matter is also challenging because the couplings
$h\chi^0_1\chi^0_1$ and $Z\chi^0_1\chi^0_1$ are suppressed by a small
factor $m_W/M_{1,2}$ for a higgsino-like $\chi^0_1$.
The spin-dependent cross section with proton due to the $Z\chi^0_1\chi^0_1$
coupling is approximately given by
\cite{Hisano:2004pv}
\begin{align}
\sigma_{\rm SD} \sim 0.8\times10^{-42}{\rm cm}^2
\times \left(\frac{M_2}{10^4\GeV}\right)^{-2}
\left(\frac{\mu}{100\GeV}\right)^{-2}\cos^22\beta,
\end{align}
for $M_1=M_2$.
The spin-independent scattering is mediated mainly by Higgs boson exchange,
and has a cross section smaller than $\sigma_{\rm SD}$ by about 5 orders
of magnitude.
Thus, in both cases, the scattering is too small to be detectable by current
experiments.
On the other hand, the higgsino annihilation into $\gamma\gamma$ and $\gamma Z$
can provide an interesting signal for the indirect detection of dark matter.
Dark matter is detectable by observing a $\gamma$-ray line with energy
$E_\gamma=m_{\chi^0_1}$ or $m_{\chi^0_1} - m^2_Z/4m_{\chi^0_1}$
coming from the Galatic center.
The cross section for $\chi^0_1\chi^0_1\to\gamma\gamma$ is
$\sigma v\approx 10^{-28}{\rm cm}^3/{\rm s}$ for the higgsino LSP, and the
annihilation into $\gamma Z$ has a little bit larger cross section
\cite{Bergstrom:1997fh,Ullio:1997ke}.
The annihilation cross sections are about one order of magnitude below the
experimental upper limits \cite{Abdo:2010dk}.
Future experiments would be possible to observe $\gamma$-ray lines
from the processes $\chi^0_1\chi^0_1\to\gamma\gamma,\gamma Z$, providing
a signature of dark matter.

Though we have focused on the case with $\mu\ll m_{\rm soft}\sim m_{3/2}/8\pi^2$,
it would be possible to have a light gaugino with mass $\ll m_{3/2}/8\pi^2$
if the anomaly mediation contribution is cancelled by some other contribution.
Then mass differences among the light charginos and neutralinos become larger
as long as the higgsino remains the dominant component of $\chi^0_1$.
This increases the detection potential of SUSY at the LHC
as the decay of chargino can produce hard enough leptons or jets.
In addition, because the couplings $h\chi^0_1\chi^0_1$ and $Z\chi^0_1\chi^0_1$
are enhanced, there are more possibilities also for the direct detection
of dark matter.

To conclude, we have shown that the heavy gravitino scenario with
the hierarchical mass spectrum
\begin{align}
\mu \sim 10^2\:\GeV \ll
m_{\rm soft}\sim 10^4\:\GeV
\ll m_{3/2}\sim 10^6\:\GeV
\end{align}
can explain well the EWSB while solving the cosmological problems
associated with the gravitino.
The EWSB is achieved at the correct scale for
$B\sim m_{3/2} \sim 8\pi^2 m_{\rm soft}$, as is generally the case when
the anomaly mediation is a main source of superparticle masses.
Furthermore, the higgsino LSP abundance produced by the gravitino decays
is consistent with the observed value.
We also note that the heavy stop with mass $m_{\rm soft}\sim 10^4\:\GeV$
yields the SM-like Higgs boson mass around 125 GeV without invoking
a large stop mixing.

\acknowledgments

We would like to thank Ryuichiro Kitano for useful discussion.
This work was supported by Grants-in-Aid for Scientific Research from
the Ministry of Education, Science, Sports, and Culture (MEXT),
Japan, No. 23104008 and No. 23540283.

\end{document}